\documentclass[10pt,twocolumn,showpacs,longbibliography,prb,aps,floatfix]{revtex4-1}

\usepackage{graphicx}
\usepackage{amsmath}
\usepackage{amssymb}
\usepackage{bm}
\usepackage{mathrsfs}

\newcommand{\Ham}{\hat{\mathcal{H}}}
\renewcommand{\Re}{\text{Re }}
\renewcommand{\Im}{\text{Im }}

\newcommand{\eps}{\varepsilon}
\newcommand{\om}{\omega}
\newcommand{\lam}{\lambda}
\newcommand{\Th}{\Theta}

\begin{document}

\title{Frequency-dependent polarizability, plasmons, and screening in the 2D pseudospin-1 dice lattice}

\author{J.D. Malcolm}
\author{E.J. Nicol}
\affiliation{Department of Physics, University of Guelph, Guelph, Ontario N1G 2W1 Canada} 
\affiliation{Guelph-Waterloo Physics Institute, University of Guelph, Guelph, Ontario N1G 2W1 Canada}
\date{\today}

\begin{abstract}
{We calculate the dynamic polarizability under the random phase approximation for the dice lattice.  This two-dimensional system gives rise to massless Dirac fermions with pseudospin-1 in the low-energy quantum excitation spectrum, providing a Dirac-cone plus flat-band dispersion.  Due to the presence of the flat band, the polarizability shows key differences to that of graphene (the pseudospin-1/2 Dirac material).  We find that the plasmon branch is pinched in to a single point, $\om_p=q=\mu$, independent of the background dielectric constant.  Finally, screening effects are discussed with regard to impurities.}
\end{abstract}



\maketitle

\section{Introduction}

\begin{figure}
\begin{center}
\includegraphics[width=1.0\linewidth]{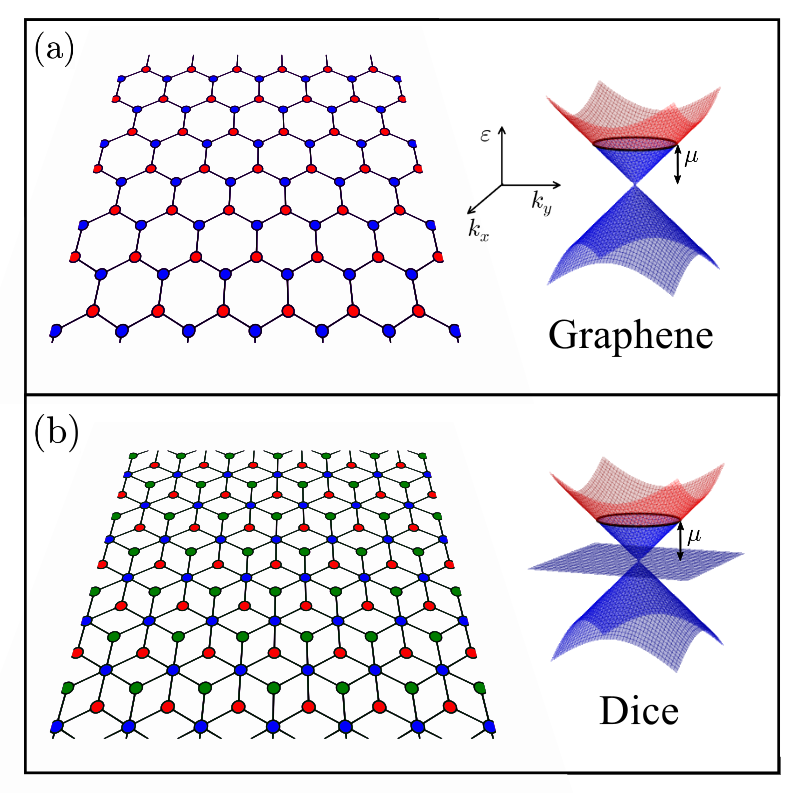}
\end{center}
\caption{\label{fig:Lattice}(Color online) (a)~Graphene's honeycomb lattice, constructed from two triangular sublattices (red and blue sites), presented with the low-energy quasiparticle dispersion.  (b)~The dice lattice, composed of three triangular sublattices (red, blue, and green sites), also with its low-energy quasiparticle dispersion.  In both diagrams, the chemical potential $\mu$ is indicated, marking the separation between occupied (blue) and unoccupied (red) states.}
\end{figure}

Graphene, discovered in 2004\cite{Novoselov04}, is the hallmark Dirac material, a material whose quantum excitations are described by Dirac physics\cite{Novoselov05,Zhang05}.  Restricted to two dimensions (2D), this honeycomb lattice of carbon atoms (Fig.~\ref{fig:Lattice}(a)) has been an intense subject of research at both the fundamental and applied levels (see various reviews\cite{Geim07,Zhu10,DasSarma11,Kotov12}).  Many of the exotic properties of graphene are a result of the low-energy dynamics of charge carriers, whose behaviour is described by the massless Dirac Hamiltonian in 2D,
\begin{equation}\label{eqn:HamGraphene}
\Ham_g = \hbar v
\begin{pmatrix}
0 & k_- \\
k_+ & 0 \\
\end{pmatrix} = \hbar v\,\bm{\sigma}\cdot\bm{k}\,,
\end{equation}
where $v$ is the Fermi velocity and $\bm{\sigma} = (\sigma_x,\sigma_y)$ are the first two Pauli matrices which are contracted with the 2D wavevector $\bm{k}=(k_x,k_y)$, with $k_\pm=k_x\pm ik_y$.  The subscript on $\Ham_g$ refers to graphene specifically.  In the special case of 2D, the massless Dirac equation is exactly the chiral Weyl equation.  Eq.~(\ref{eqn:HamGraphene}) describes the Dirac fermions in graphene around one chiral centre, or valley, in the reciprocal lattice, called the K point, while another set of fermions exist at the K$'$ point with opposite chirality ($\bm{\sigma}\rightarrow-\bm{\sigma}^*$).  The Hamiltonian in Eq.~(\ref{eqn:HamGraphene}) gives the so-called Dirac-cone energy dispersion (Fig.~\ref{fig:Lattice}(a)), where particles and holes have linear-in-momentum energy dependence with a single-point band crossing at zero energy. 

Recognizing the Pauli matrices as being equal to the spin-1/2 matrices (up to a factor of 2), a simple generalization of Eq.~(\ref{eqn:HamGraphene}) is to use instead the spin-$s$ matrices,
\begin{equation}\label{eqn:HamDW}
\Ham_{DW} = \hbar v_c\,\bm{S}_s\cdot\bm{k}\,,
\end{equation}
giving the generalized 2D Dirac-Weyl Hamiltonian of pseudospin-$s$, with $v_c$ being a factor with dimensions of velocity that is characteristic to the system being studied.  As such, graphene is specifically characterized as a pseudospin-1/2 Dirac-Weyl material (whose Fermi velocity $v=v_c/2$), with a 2-spinor as the single-particle wavefunction.  The term pseudospin is used because it refers to an emergent $\rm SU(2)$ symmetry that exists in addition to the intrinsic electron spin.  In graphene, the pseudospin is an index for the two triangular sublattices (blue and red in Fig. \ref{fig:Lattice}(a)) and does not describe the quantization of a magnetic moment (as it does for intrinsic spin).  However, the pseudospin still does have an angular momentum associated with it\cite{Mecklenburg11}.  

The main subject of this study is the 2D dice lattice\cite{Horiguchi74} (also called the $\mathcal{T}_3$ lattice\cite{Vidal98}), shown in Fig.~\ref{fig:Lattice}(b).  With three triangular sublattices, as opposed to graphene's two, the single-particle wavefunction in this Dirac-Weyl system is a 3-spinor\cite{Dora11}.  Hence, the low-energy dispersion around one chiral centre (of which there are two in the dice lattice) is described by the pseudospin-$1$ Hamiltonian\cite{Bercioux09},
\begin{equation}\label{eqn:HamDice}
\Ham_d = \frac{\hbar v}{\sqrt{2}}
\begin{pmatrix}
0 & k_- & 0 \\
k_+ & 0 & k_- \\
0 & k_+ & 0 \\
\end{pmatrix}\,,
\end{equation}
where, in comparision to Eq.~(\ref{eqn:HamDW}), the Fermi velocity is $v=v_c$.  Of significant interest, the energy dispersion from Eq.~(\ref{eqn:HamDice}) (shown in Fig.~\ref{fig:Lattice}(b)) exhibits an entirely flat band at zero energy for all momentum, in addition to the Dirac cones found in graphene.  This dispersionless band results from the conformity of a three-band system with the particle-hole symmetry of Dirac fermions.  The large degeneracy found in this flat band gives rise to a singular density of states, which can have a large impact on various physical features of a material\cite{Heikkila11,Louvet15,Illes15}.

The property we are interested in here is the dynamical polarizability, $Q(\bm{q},\om)$, an entity from many-body physics which renormalizes the Coulomb interaction between charge carriers\cite{Mahan13}.  This renormalization is achieved through modification of the dielectric function,
\begin{equation}\label{eqn:Dielectric}
\eps(\bm{q},\om) = 1 + V_c(q)Q(\bm{q},\om)\,,
\end{equation}
where, in 2D, the bare Coulomb interaction is $V_c(q) = 2\pi\alpha/q$ and the Wigner-Seitz radius, $\alpha=e^2/(\epsilon_0\hbar v)$, plays the role of an effective fine structure constant\cite{Hwang07,DasSarma09} in which $\epsilon_0$ is the background dielectric constant (i.e., the substrate dielectric).  For physical graphene on a $\rm SiO_2$ substrate, $\alpha\sim0.5$.

Under the random-phase approximation (RPA), we have derived an analytic form for the low-energy dice lattice polarizability, which includes the Lindhard function in the static ($\om=0$) regime.  As discussed below, the results show similar features to graphene's polarizability with additional aspects arising from the flat band's presence.  Using this to recast the dielectric function, we are able to make a numerical analysis of plasmon oscillations and screening within the material, both effects showing marked differences from those in graphene.

While a material constructed from the dice lattice has not yet been observed to exist naturally, the system in Eq.~(\ref{eqn:HamDice}) could be constructed artificially using ultra-cold gases of atoms or with photonic lattices\cite{Polini13}.  Further, based on a set of optical measurements\cite{Orlita14}, we have previously discussed the pseudospin qualities of the three-dimensional material ${\rm Hg}_{1-x}{\rm Cd}_x{\rm Te}$ (MCT) under the critical fraction $x=x_c\approx0.17$\cite{Malcolm15}.  Confining MCT to 2D, the low-energy Hamiltonian maps onto a specific case ($\alpha=1/\sqrt{3}\,$) of the $\alpha$-$\mathcal{T}_3$ model\cite{Raoux14} (where this $\alpha$ is different from the fine-structure constant described above).  The $\alpha$-$\mathcal{T}_3$ model interpolates between the graphene lattice (Fig.~\ref{fig:Lattice}(a)) and the dice lattice (Fig.~\ref{fig:Lattice}(b)) as $\alpha$ varies continuously from $0$ to $1$.  This interpolation employs a variable hopping parameter associated with the third triangular sublattice present in the dice lattice, but not in graphene (green sites in Fig.~\ref{fig:Lattice}(b)).  Thus, MCT might be considered as a Dirac material existing as a hybrid of pseudospin-$1/2$ and $1$.  The opportunity of studying physical materials with pseudospin-$1$ characteristics motivates the need for a theoretical analysis of this system's properties.

In the remainder of the paper, we have taken the liberty of assigning $\hbar=v=1$ for simplicity.  In addition, by restricting the chemical potential, $\mu$, to finite values only, partial filling of the flat band is not considered.  Throughout, it is sufficient to consider only positive values of $\mu$ due to the particle-hole symmetry of Dirac fermions.

\section{Dynamical Polarizability}

In the Feynman diagram representation, the RPA takes the polarizability to be a particle-hole bubble.  Mathematically, this is
\begin{equation}\label{eqn:Bubble}
Q(\bm{q},\om) = \frac{g}{4\pi^2}\int d^2\bm{k}\sum_{\lam,\lam'}\frac{f_{\lam'\bm{k}'}-f_{\lam\bm{k}}}{\om-(\eps_{\lam'\bm{k}'}-\eps_{\lam\bm{k}})+i\eta}F_{\lam\lam'}(\bm{k})\,,
\end{equation}
where $g$ is a degeneracy factor (like in graphene, $g=4$ for the dice lattice due to twofold valley and intrinsic-spin degeneracies), $\lam$ is a band index, $\eps_{\lam\bm{k}}$ are the energy eigenvalues, $\eta=0^+$ can be viewed as an infinitesimal scattering rate, $\bm{k}'=\bm{k}+\bm{q}$, and $f_{\lam\pmb{k}}$ is the Fermi-Dirac distribution under chemical potential $\mu$ for the energy $\eps_{\lam\bm{k}}$.  The numerator containing the two statistical functions ensures that the integrand covers only the overlap of particle-hole pairs $|\lam\bm{k}\rangle$ and $|\lam'\bm{k}'\rangle$ (as opposed to particle-particle or hole-hole pairs).  The definition of the overlap function is
\begin{equation}
F_{\lam\lam'} = F_{\lam'\lam} = |\langle\lam\bm{k}|\lam'\bm{k}'\rangle|^2\,.
\end{equation}

\begin{figure}[h!]
\begin{center}
\includegraphics[width=1.0\linewidth]{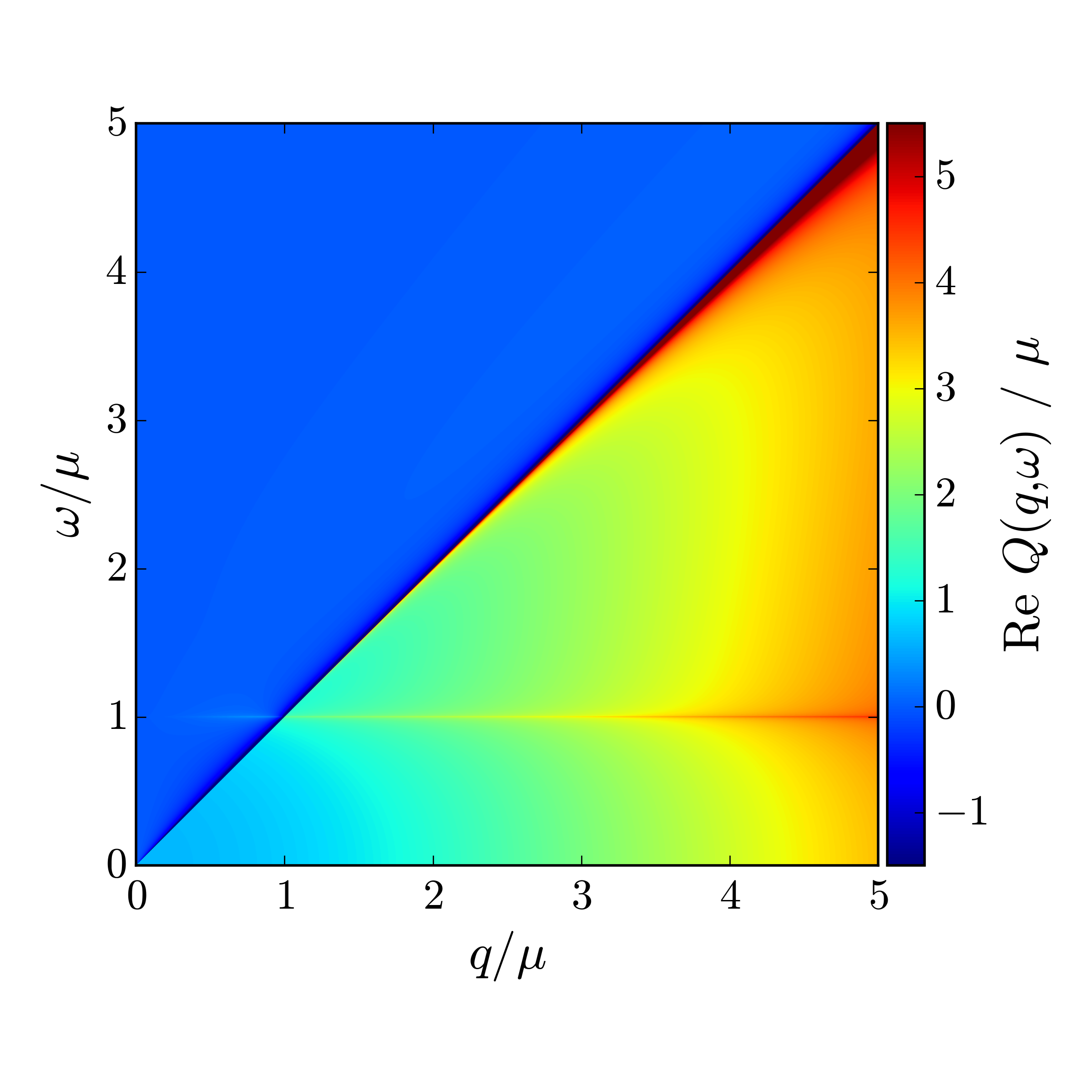}
\end{center}
\caption{\label{fig:ReQ}(Color online) Real part of the dice lattice polarization function, Eq.~(\ref{eqn:ReQ}).}
\end{figure}

\begin{figure}[t!]
\begin{center}
\includegraphics[width=1.0\linewidth]{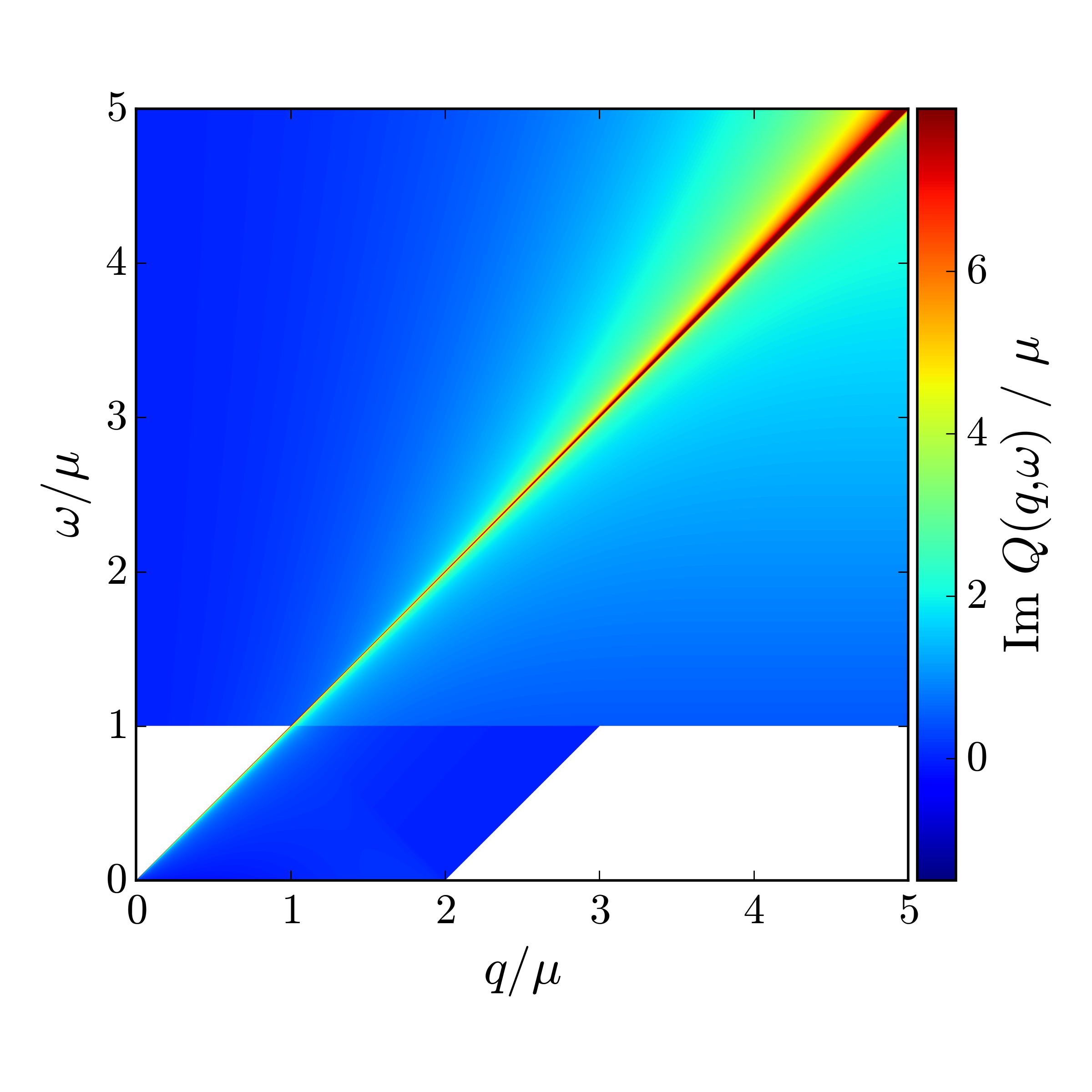}
\end{center}
\caption{\label{fig:ImQ}(Color online) Imaginary part of the dice lattice polarization function, Eq.~(\ref{eqn:ImQ}).  White regions mark voids in the particle-hole continuum, or regions where the imaginary part is exactly zero.}
\end{figure}

From the low-energy dice-lattice Hamiltonian in Eq.~(\ref{eqn:HamDice}): $\lam=\{0,\pm1\}$ and $\eps_{\lam\bm{k}}=\lam k$.  With $\mu>0$, only the $F_{\lam+}$ matrix elements are required,
\begin{subequations}
\begin{align}
F_{0+}(\bm{k}) &= \frac{1}{2}\sin^2\phi_{\bm{k}}\,, \\
\label{eqn:Overlapb}F_{\pm+}(\bm{k}) &= \frac{1}{4}(1\pm\cos\phi_{\bm{k}})^2\,,
\end{align}
\end{subequations}
where $\phi_{\bm{k}}$ is the angle between $\bm{k}=(k,\theta)$ and $\bm{k}'$.  That is, for $k_x$ aligned with $\bm{q}$,
\begin{equation}
\cos\phi_{\bm{k}} = \frac{\bm{k}\cdot\bm{k}'}{kk'} = \frac{k+q\cos\theta}{\sqrt{k^2+q^2+2qk\cos\theta}}\,.
\end{equation}
Eq.~(\ref{eqn:Bubble}) has been solved for doped graphene\cite{Shung86,Ando06,Wunsch06,Hwang07} as well as under additional considerations of gapped graphene\cite{Pyatkovskiy09} and lattice buckling\cite{Tabert14}.  By the same procedure, we have calculated the result of Eq.~(\ref{eqn:Bubble}) for the dice lattice at zero temperature.  The zero-temperature assumption reduces the Fermi-Dirac distribution to $f_{-\bm{k}}=f_{0\bm{k}} = 1$ for the lower cone and flat band, and $f_{+\bm{k}}=\Th(\mu-k)$ for the upper cone, making use of the Heaviside step function, $\Th(x)$.  Subsequent application of the identity $Q(\bm{q},\om)=-Q^*(\bm{q},-\om)$ allows for the consideration of $\om>0$ only.  We present the result here, first defining
\begin{equation}\label{eqn:hFunctions}
\begin{split}
& h_1(x) = -x\,\sqrt{\left|\frac{x^2-q^2}{\om^2-q^2}\right|}\,, \\
& h_{2A}(x) = \frac{3q^2-2\om^2}{\sqrt{|\om^2-q^2|}}\arccos\left(\frac{x}{q}\right)\,, \\
& h_{2B}(x) = \frac{3q^2-2\om^2}{\sqrt{|\om^2-q^2|}}\ell n\left|\frac{x+\sqrt{|x^2-q^2|}}{q}\right|\,, \\
& h_{3A}(x) = \frac{\om^2-q^2}{\om}\ell n\left|\frac{\om\sqrt{|x^2-q^2|}+x\sqrt{|\om^2-q^2|}}{\om\sqrt{|x^2-q^2|}-x\sqrt{|\om^2-q^2|}}\right|\,, \\
& h_{3B}(x) = 2\frac{\om^2-q^2}{\om}\arccos\left(\frac{x}{q}\sqrt{\left|\frac{\om^2-q^2}{\om^2-x^2}\right|}\right)\,.
\end{split}
\end{equation}
The real part of the dice lattice polarizability is
\begin{equation}\label{eqn:ReQ}
\begin{split}
&\left(\vphantom{\frac{x}{x}}\Re Q(\bm{q},\om)\right)/\frac{g}{16\pi} \\
& = 8\mu+\frac{\om^2+q^2}{\om}\ell n\left|\frac{\om+\mu}{\om-\mu}\right| \\
& +\Th(q-\om)\Th(q-x)\!\!\left.\left[\vphantom{\int}h_1(x)+h_{2A}(x)+h_{3A}(x)\right]\!\right|_{x=2\mu-\om} \\
& +\Th(q-x)\!\!\left.\left[\vphantom{\int}h_1(x)+h_{2A}(x)+h_{3A}(x)\right]\!\right|_{x=2\mu+\om} \\
& -\Th(\om-q)\Th(x-q)\!\!\left.\left[\vphantom{\int}h_1(x)+h_{2B}(x)+h_{3A}(x)\right]\!\right|_{x=|2\mu-\om|} \\
& +\Th(\om-q)\!\!\left.\left[\vphantom{\int}h_1(x)+h_{2B}(x)+h_{3A}(x)\right]\!\right|_{x=2\mu+\om}\,,
\end{split}
\end{equation}
with the ($\om>0$) imaginary part,
\begin{equation}\label{eqn:ImQ}
\begin{split}
&\left(\Im Q(\bm{q},\om)\vphantom{\frac{x}{x}}\right)/\frac{g}{16\pi} \\
& = \!\Th(\om-\mu)\left[\vphantom{\int}2\pi\frac{\min(\om^2,q^2)}{\om}\right] \\
& +\Th(q-\om)\Th(x-q)\!\!\left.\left[\vphantom{\int}h_1(x)\!+\!h_{2B}(x)\!+\!h_{3B}(x)\right]\!\right|_{x=2\mu-\om} \\
& -\!\Th(q-\om)\Th(x-q)\!\!\left.\left[\vphantom{\int}h_1(x)\!+\!h_{2B}(x)\!+\!h_{3B}(x)\right]\!\right|_{x=2\mu+\om} \\
& +\!\Th(\om-q)\Th(q-|x|)\!\!\left.\left[\vphantom{\int}h_1(x)\!+\!h_{2A}(x)\!+\!h_{3B}(x)\right]\!\right|_{x=2\mu-\om} \\
& +\!\Th(\om-q-2\mu)\left[\vphantom{\int}\pi\frac{3q^2-2\om^2}{\sqrt{\om^2-q^2}}\!+\!2\pi\frac{\om^2-q^2}{\om}\right]\,.
\end{split}
\end{equation}
Eqs.~(\ref{eqn:ReQ}) and (\ref{eqn:ImQ}) are the main result of this work and are presented as colour-map plots in Figs.~(\ref{fig:ReQ}) and (\ref{fig:ImQ}).

The flat band's presence in the dice model produces two effects that make its polarizability different from that of graphene.  The first is a direct effect whereby extra terms arise due to the tracing over the flat band in Eq.~(\ref{eqn:Bubble}).  Second, the presence of the flat band indirectly affects cone-to-cone scattering because of the larger Hilbert space (i.e., the dice-lattice 3-spinor compared to graphene's 2-spinor).  This is manifest in the cone-to-cone scattering amplitude (Eq.~(\ref{eqn:Overlapb})) when compared to the same in graphene which is merely proportional to $(1\pm\cos\phi_{\bm{k}})$ and not its square.

Several straight-forward manipulations on our result give the polarizability of graphene.  Referring to Eqs.~(\ref{eqn:hFunctions}), the $h_2$ functions must be adjusted so that the $3q^2-2\om^2$ factor is simply replaced by $q^2$.  In addition, the $h_3$ functions must be removed entirely.  Further, the first line in the real part (Eq.~(\ref{eqn:ReQ})) must be replaced by the single term $8\mu$.  In the imaginary part (Eq.~(\ref{eqn:ImQ})), the first line is omitted and the behaviour in the last line is replaced by $\pi q^2/\om$.  The adjustments on Eq.~(\ref{eqn:hFunctions}) make little difference to the overall profile of the polarization.  The major distinctions found between the dice lattice and graphene are those extra terms found in the first lines of Eqs.~(\ref{eqn:ReQ}) and (\ref{eqn:ImQ}).

For $\om<\mu$, the imaginary part (Fig.~\ref{fig:ImQ}) appears similar to that of graphene, with nonzero values bound by $\om<q<\om+2\mu$.  However, a major difference occurs with a discontinuous step up at $\om=\mu$, which is purely a result of scattering from the flat band.  This step up in the imaginary part corresponds to a logarithmic divergence at $\om=\mu$ in the Kramers-Kronig-related real part plotted in Fig.~\ref{fig:ReQ}.

\begin{figure}
\begin{center}
\includegraphics[width=1.0\linewidth]{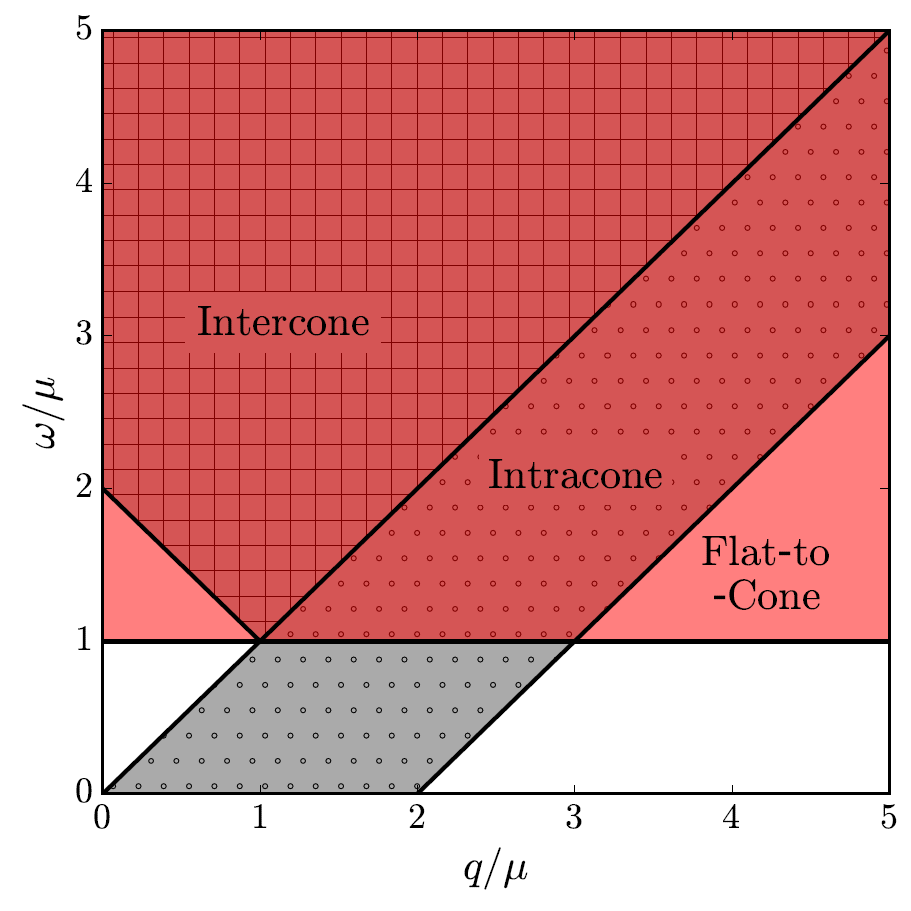}
\end{center}
\caption{\label{fig:Continuum}(Color online) The particle-hole continuum for the dice lattice in energy-momentum space.  Contributory regions from intracone (dots) and intercone (grid) scattering are shaded grey and are overlayed by the flat-to-cone contribution in red.}
\end{figure}

Non-zero regions of the imaginary part, Eq.~(\ref{eqn:ImQ}), map out the particle-hole continuum, or single-particle excitation region (Fig.~\ref{fig:Continuum}).  This is the area of $(q,\om)$-space demonstrating a geometric connection between occupied and unoccupied single-particle states separated by a specific momentum and energy.  This connection allows for the scattering of a particle into an unoccupied state via a perturbation of momentum $\bm{q}$ and energy $\om$.  In this model, where all unoccupied states are in the upper cone, the occupied states exist in all three bands: upper cone, lower cone, and flat band.  These subsets provide intracone, intercone, and flat-to-cone contributions to the different regions of the continuum, respectively, as indicated in Fig.~\ref{fig:Continuum}.  The former two contributions make up the entire particle-hole continuum in graphene.  In the dice model, the flat band greatly extends this continuum to exist for all energies that are larger than the chemical potential. This arises from the band-structure geometry (Fig.~\ref{fig:Lattice}(b)), in which for all $\om>\mu$ any value of momentum can form a connection between a state in the flat band and an unoccupied one.

\section{Dielectric Loss and Plasmons}

\begin{figure}
\begin{center}
\includegraphics[width=1.0\linewidth]{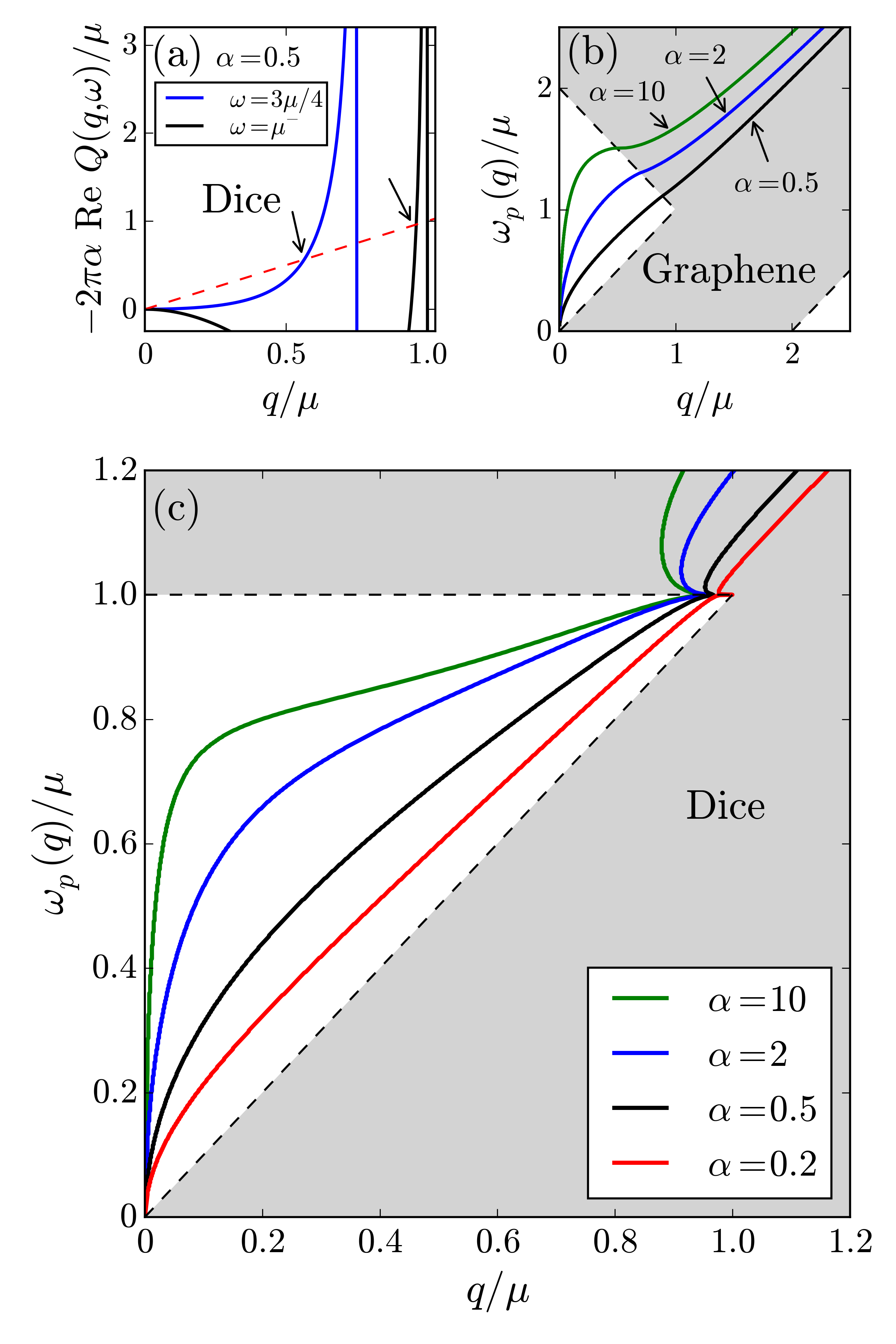}
\end{center}
\caption{\label{fig:Plasmons}(Color online) (a) An example in the construction of the dice lattice plasmon branch by graphical solution of Eq.~(\ref{eqn:Plasmon}).  Black and blue show cuts of the real part of the polarization, weighted by a particular value of $\alpha$.  In red is the line $y=q$.  Intersections of the red dashed line with the other curves are indicated. (b) Plasmon branches in graphene for different values of $\alpha$.  (c) Plasmon branches in the dice lattice for different values of $\alpha$. Note that in the red curve $\alpha$ is slightly less than $0.5$.  In both these latter plots, the particle-hole continuum is shaded in grey.}
\end{figure}

\begin{figure}
\begin{center}
\includegraphics[width=1.0\linewidth]{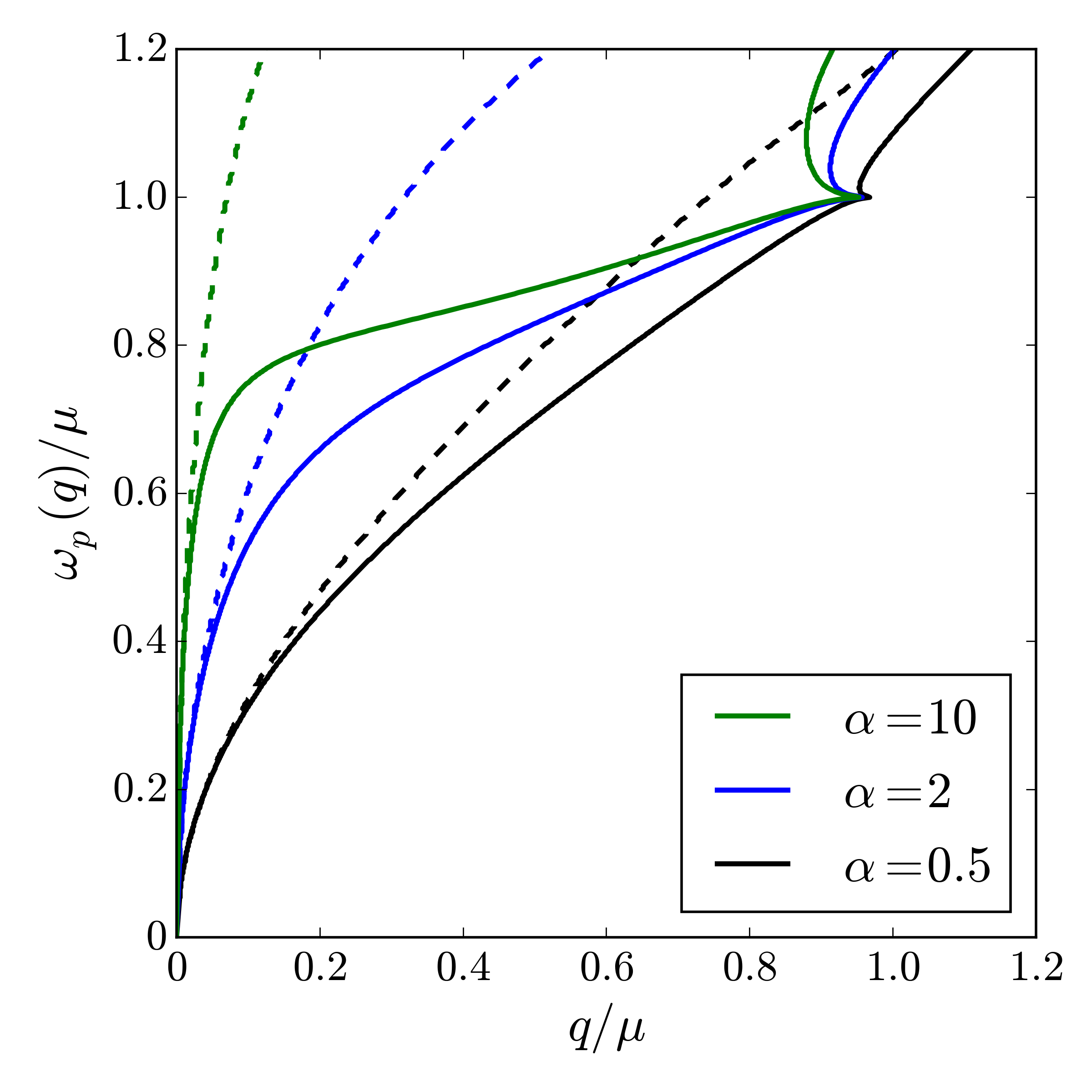}
\end{center}
\caption{\label{fig:Compare}(Color online) Dice (solid) and graphene (dashed) plasmon curves for various $\alpha$ determined through the solution of Eq.~(\ref{eqn:Plasmon}).}
\end{figure}

The energy dispersion for plasmons (collective oscillations of electrons) is determined by zeros of the dielectric function, $\eps(\bm{q},\om_p-i\gamma)=0$, where the plasmon decay rate is
\begin{equation}
\gamma(\bm{q},\om_p) = \frac{\Im Q(\bm{q},\om_p)}{\frac{\partial}{\partial\om}\Re Q(\bm{q},\om_p)}\,.
\end{equation}
Within the particle-hole continuum, plasmonic oscillations are highly damped by the creation of particle-hole pairs.  As such, we are only interested in the plasmon branch in the void beyond this continuum, where $\Im Q=0$ and thus $\gamma=0$.  Then, according to Eq.~(\ref{eqn:Dielectric}), for weak damping the plasmon branch is found by solving
\begin{equation}\label{eqn:Plasmon}
q+2\pi\alpha\Re Q(\bm{q},\om_p)=0\,.
\end{equation}
This equation has been solved graphically by varying $\om$ and determining the intersection between curves $y_1(q)=q$ and $y_2(q)=-2\pi\alpha\Re Q(\bm{q},\om)$.  This procedure is illustrated in Fig.~\ref{fig:Plasmons}(a) for $\alpha=0.5$ and two example curves of $y_2$ at $\om=3\mu/4$ and $\om=0.999\mu\equiv\mu^-$.  $y_1$ is shown in red, whose intersection with the black and blue curves marks values $\om_p(\bm{q})$  which construct the plasmon branch.  The plasmon branch is plotted in Fig.~\ref{fig:Plasmons}(c) for different values of $\alpha$ with the particle-hole continuum shaded grey.  The variation in $\alpha\sim1/\epsilon_0$ corresponds to a different background dielectric constant determined by the substrate.  An analogous picture of plasmons in graphene is provided in Fig.~\ref{fig:Plasmons}(b) for comparison.  Note that the branches are plotted in the particle-hole continuum here, but any actual plasmon oscillation would be damped out in this region.  For a more direct comparison, the plasmon curves for both 2D systems are provided on the same plot in Fig.~\ref{fig:Compare}.  This last plot assumes that the Fermi velocity, $v$, is the same in both systems, which need not be the case.

Fig.~\ref{fig:Compare} demonstrates that for low energy and momentum, plasmons behave the same in both graphene and in the dice lattice.  In this region, plasmon behaviour comes from particles near the Fermi level so that the flat band has little influence.  However, as the plasmon energy increases, heavy screening from the flat band pinches the branch in to the point $\om_p=q=\mu$, regardless of the value of $\alpha$.  Seen in Fig.~\ref{fig:Plasmons}(a), the divergence in $\Re Q$ at $\om=q$ becomes increasingly narrow as $\om$ tends to $\mu$.  As this divergence becomes infinitesimally narrow, the intersection of the red dashed line occurs essentially at $\om=\mu$, giving rise to the pinch point in the dispersion.

\begin{figure}
\begin{center}
\includegraphics[width=1.0\linewidth]{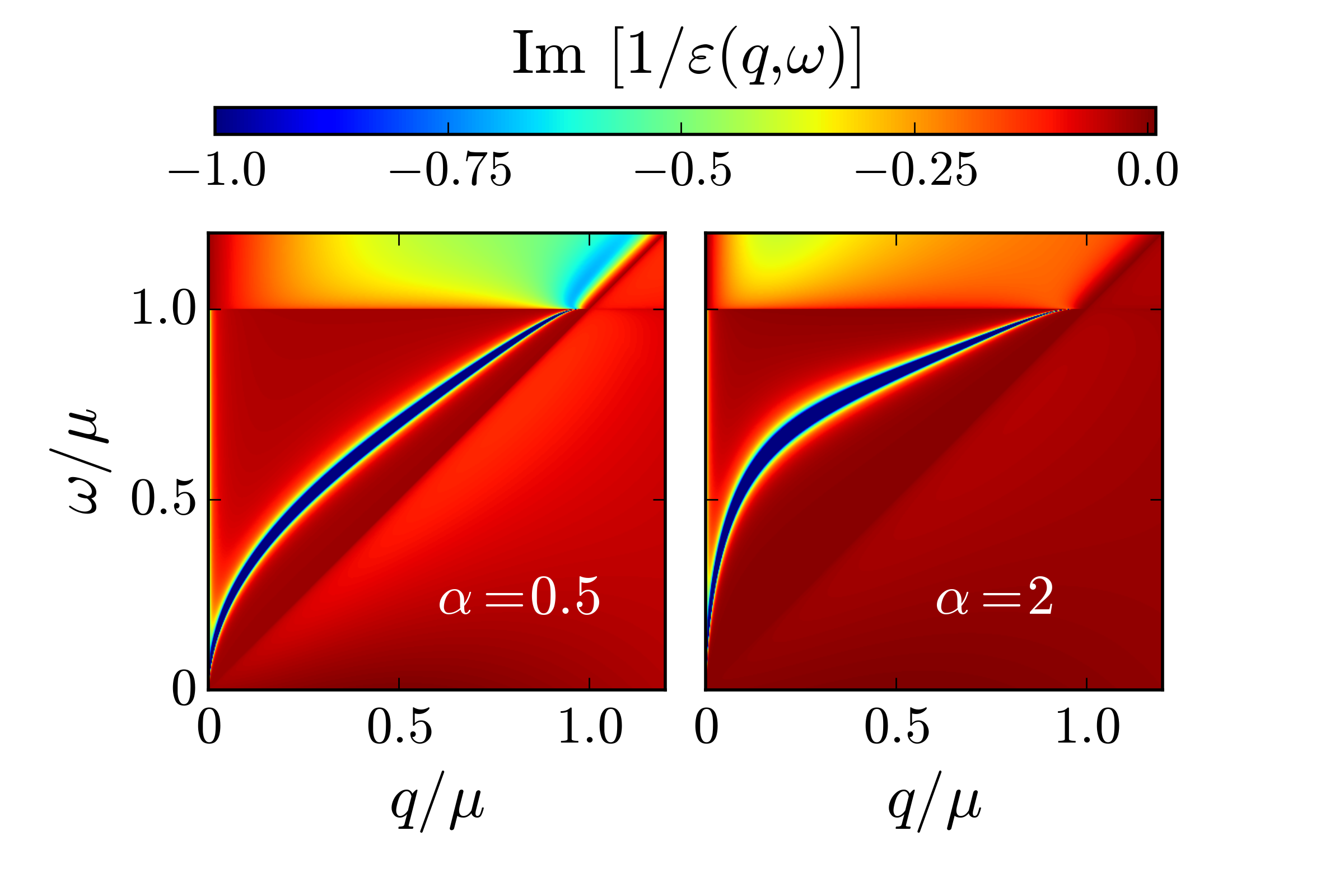}
\end{center}
\caption{\label{fig:Loss}(Color online) Dielectric loss function in the dice lattice for values $\alpha=0.5$ and $\alpha=10$.  In both plots, blue traces out a singular path through the void in the particle-hole continuum that marks the plasmon branch.}
\end{figure}

The plasmon dispersion can also be visualized through the dielectric loss function, $\Im[1/\eps(\bm{q},\om)]$, which is singular along the plasmon branch.  This is seen in Fig.~\ref{fig:Loss} for example values of $\alpha=0.5$ and $\alpha=10$, with the plasmon branch traced out in blue in the particle-hole void in each plot.  These plots were constructed using Eqs.~(\ref{eqn:Dielectric}), (\ref{eqn:ReQ}), and (\ref{eqn:ImQ}).  Note that the undamped branches exist only outside the particle-hole continuum where $q<\om<\mu$ and otherwise match their counterparts in Fig.~\ref{fig:Plasmons}(c).

\section{The Lindhard Function and Static Screening Effects}

\begin{figure}
\begin{center}
\includegraphics[width=1.0\linewidth]{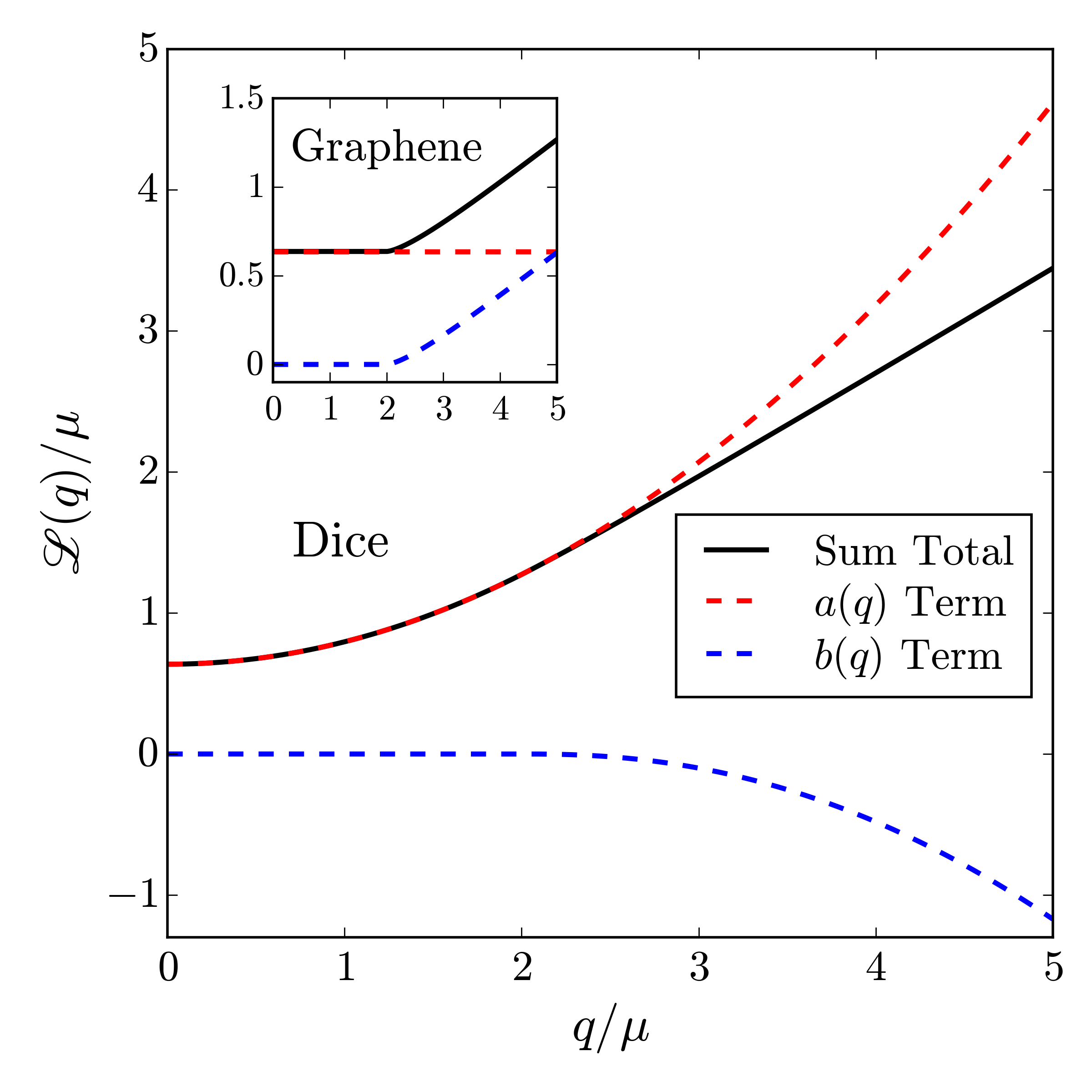}
\end{center}
\caption{\label{fig:Lindhard}(Color online) Dice lattice Lindhard function with constituent $a$- and $b$-terms in red and blue, respectively (see Eq.~(\ref{eqn:LindhardDice})).  Inset is the same for graphene (Eq.~(\ref{eqn:LindhardGraphene})).}
\end{figure}

The zero-frequency limit of the polarization function gives the static Lindhard function, $\mathscr{L}(\bm{q})=\Re Q(\bm{q},\om=0)$, henceforth referred to as the Lindhard function.  For graphene and the dice lattice, this takes the form
\begin{equation}
\mathscr{L}(\bm{q}) = a(q)+\Th(q-2\mu)b(q)\,.
\end{equation}
The $a$-term persists for all $q$, whereas the $b$ term is only onset for $q>2\mu$.  The 2D electron gas (2DEG) also has a Lindhard function of this form\cite{Stern67}.  However, one must replace $\mu$ with $k_F$, since these quantities are not equal in the 2DEG.  Explicitly, for graphene\cite{Kotov12}
\begin{equation}\label{eqn:LindhardGraphene}
\begin{split}
a_g(q) &= \frac{g}{2\pi}\mu\,, \\
b_g(q) &= \frac{g}{8\pi}\left[q\arccos\left(\frac{2\mu}{q}\right)-\frac{2\mu}{q}\sqrt{q^2-4\mu^2}\right]\,;
\end{split}
\end{equation}
and for the dice lattice, we find that
\begin{equation}\label{eqn:LindhardDice}
\begin{split}
a_d(q) &= \frac{g}{8\pi}\left(4\mu+\frac{q^2}{\mu}\right) \\
b_d(q) &= \frac{g}{8\pi}\left[3q\arccos\left(\frac{2\mu}{q}\right)-\left(\frac{2\mu}{q}+\frac{q}{\mu}\right)\sqrt{q^2-4\mu^2}\right]\,.
\end{split}
\end{equation}
The dice-lattice Lindhard function is plotted in Fig.~\ref{fig:Lindhard} with its constituent $a$- and $b$-terms.  Inset in this figure is the same for graphene.  Note that for graphene, the $a$-term is a constant, proportional to the chemical potential $\mu$.  This then leaves a plateau in the Lindhard function for $q<2\mu$.   The $a$-term in the dice lattice, however, is parabolic.  Beyond $q=2\mu$, the $b$-term in graphene is positive, leading to an increase in the function.  In the dice lattice, this term is negative, only slightly correcting the parabolic $a$-term.  Importantly, the graphene Lindhard function is most singular in its second derivative, while for the dice lattice this behaviour is found in the third derivative.

From the Lindhard function, it is possible to determine the response of the Dirac fermions in the material to the inclusion of an electric or magnetic impurity.  The potential around a point charge, $Q$, is proportional to the Fourier transform of the renormalized Coulomb interaction,
\begin{equation}\label{eqn:phi}
\begin{split}
\varphi(\bm{r}) &= \frac{Q}{\alpha\epsilon_0}\int\frac{d^2\bm{q}}{(2\pi)^2}\frac{V_c(q)}{\eps(q)}e^{-i\bm{q}\cdot\bm{r}} \\
&= \frac{Q}{\epsilon_0}\int_0^\infty dq\frac{J_0(qr)}{\eps(q)}\,,
\end{split}
\end{equation}
where, in the static limit, $\eps(q)=1+V_c(q)\mathscr{L}(q)$.  $J_0(x)$ is the zeroth Bessel function of the first kind.

\begin{table}[t]
\caption{\label{tab:Screening}Spatial dependence in the screening potential around a charge impurity in three different 2D systems.  The dominant behaviour is the Thomas-Fermi (TF) decay and next order is the Friedel oscillations.  Note that in the relativistic systems, $k_F=\mu$, and the two parameters can be interchanged here.}
\begin{tabular}{l c c}
System & TF & Friedel \\
\hline
2DEG & $1/r^3$ & \hspace{10pt} $\cos(2k_F r)/r^2$ \\
Graphene \hspace{10pt} & $1/r^3$ & \hspace{10pt} $\cos(2\mu r)/r^3$ \\
Dice & $1/r^3$ & \hspace{10pt} $\cos(2\mu r)/r^4$ \\
\end{tabular}
\end{table}

The dominant behaviour in $\varphi(r)$ is the Thomas-Fermi (TF) decay.  For each of the 2D systems discussed here (2DEG, graphene, and dice), the TF decay has $1/r^3$ dependence.  At next highest order, Friedel oscillations are present, arising from the singular behaviour of the Lindhard function at $q=2\mu$.  Lighthill theorem says that the asymptotic behaviour of a Fourier transform is dominated by the singular behaviour of the integrand or any of its derivatives\cite{Lighthill58}.  As noted above, the derivatives $\mathscr{L}_g^{(2)}$ and $\mathscr{L}_d^{(3)}$ have the most singular behaviour for graphene and dice, respectively.  In Eq.~(\ref{eqn:phi}), these singularities are teased out by first taking the long wavelength limit and then integrating by parts (IBP) until the desired order of derivative appears in the integrand.  Each IBP pulls out a factor of $1/r$, so that a singularity in a higher derivative leads to a stronger decay of the Friedel oscillations.  As such, the graphene Friedel oscillations have $\cos(2\mu r)/r^3$ dependence and those in the dice lattice have $\cos(2\mu r)/r^4$.  For further comparison, the 2DEG Linhard function is singular in its first derivative and its Friedel oscillations decay as $\cos(2k_F r)/r^2$.  These decay rates are all summarized in Table~{\ref{tab:Screening}.

Both the induced spin texture around a magnetic impurity and the Ruderman-Kittel-Kasuya-Yosida (RKKY) interaction strength between two magnetic impurities are proportional to the Fourier transform of the Lindhard function itself\cite{BealMonod87,Wunsch06},
\begin{equation}
\begin{split}
\mathscr{L}(\bm{r}) &= \int\frac{d^2\bm{q}}{(2\pi)^2}\mathscr{L}(q)e^{-i\bm{q}\cdot{r}} \\
&= \frac{1}{2\pi}\int_0^\infty dq\,qJ_0(qr)\mathscr{L}(q)\,.
\end{split}
\end{equation}
In this magnetic screening, there is no TF decay like for an electric impurity, but the Friedel oscillations have the same spatial dependence as presented in Table~{\ref{tab:Screening}.

\section{Conclusions}

We have obtained an analytic expression for the dice lattice polarizability, allowing for subsequent calculation of plasmon and screening behaviours in the material.  The presence of the flat band in this system, when compared to graphene, provides notable alteration of the physics, including an extended particle-hole continuum.  Due to strong screening in the flat band, the plasmon branch in the dice lattice is pinched in to the point $\om_p=q=\mu$, independent of the substrate properties.  Finally, Friedel oscillations in the screening of electric and magnetic impurities were shown to decay faster than those in the 2DEG and graphene due to the distinct singular nature of the Lindhard function.  Subsequent analyses can be built up from our work by examining, for example, polariton, plasmaron, or other collective behaviour in the dice lattice.

The study of Dirac materials remains a current and exciting field, with evermore frequent discussions of systems exhibiting pseudospin beyond $1/2$.  The polarizability of the pseudospin-1 system is an entity fundamental to the many-body physics at work in this material, allowing for the description of many of its physical properties.

We would like to thank J.P. Carbotte for a critical reading of this manuscript.  In the review process, the referees were helpful in pointing out an error in an initial calculation, for which we are appreciative.  We are also grateful to B.G. Nickel and C.J. Tabert for helpful discussions surrounding this work, which has been supported by the Natural Science and Engineering Research Council of Canada.

\bibliography{bibliography}
\end{document}